# Robust fractional quantum Hall effect and composite fermions in the $N$=2 Landau level in bilayer graphene

Georgi Diankov[1†], Chi-Te Liang[1,2†*], François Amet[3,4], Patrick Gallagher[1], Menyoung Lee[1], Andrew J. Bestwick[1], Kevin Tharratt[1], William Coniglio[5], Jan Jaroszynski[5], K. Watanabe[6], T. Taniguchi[6] and David Goldhaber-Gordon[1*]

[1]Department of Physics, Stanford University, Stanford, California 94305, USA
[2]Department of Physics, National Taiwan University, Taipei 106, Taiwan
[3]Department of Physics, Duke University, Durham, North Carolina 27708, USA
[4]Department of Physics and Astronomy, Appalachian State University, Boone, NC 28608, USA
[5]National High Magnetic Field Laboratory, Tallahassee, Florida 32310, USA
[6]Advanced Materials Laboratory, National Institute for Materials Science, 1-1 Namiki, Tsukuba 305, Japan

[†]These authors contributed equally to this work.
[*]e-mail: ctliang@phys.ntu.edu.tw or goldhaber-gordon@stanford.edu

The fractional quantum Hall[1] (FQH) effect is a canonical example of electron-electron interactions producing new ground states in many-body systems[2]. Most FQH studies have focused on the lowest Landau level (LL), whose fractional states are successfully explained by the composite fermion (CF) model, in which an even number of magnetic flux quanta are attached to an electron[3] and where states form the sequence of filling factors $\nu = p/(2mp\pm1)$, with $m$ and $p$ positive integers. In the widely-studied GaAs-based system, the CF picture is thought to become unstable for the $N\geq2$ LL (Ref. 4), where larger residual interactions between CFs are predicted and competing many-body phases have been observed[4–7]. Here we report transport measurements of FQH states in the $N$=2 LL (filling factors $4 < |\nu| < 8$) in bilayer graphene, a system with spin and valley degrees of freedom in all LLs, and an additional orbital degeneracy in the 8-fold degenerate $N$=0/$N$=1 LLs. In contrast with recent observations of particle-hole asymmetry in the $N$=0/$N$=1 LLs of bilayer graphene,[8] the FQH states we observe in the $N$=2 LL are consistent with the CF model: within a LL, they form a complete sequence of particle-hole symmetric states whose relative strength is dependent on their denominators. The FQH states in the $N$=2 LL display energy gaps of a few Kelvin, comparable to and in some cases larger than those of **fractional states in the *N*=0/*N*=1 LLs. The FQH states we observe form, to the best of our knowledge, the highest set of particle-hole symmetric pairs seen in any material system.**

A FQH state was first observed at LL filling factor $v$ = 1/3 in a GaAs/AlGaAs two-dimensional (2D) electron system[1]. This many-body state was successfully explained by the Laughlin wave-function[2]. Within the lowest LL, the $v$ = 2/3 = 1-1/3 state was interpreted as the particle-hole conjugate[9] of the $v$=1/3 "Laughlin state". The realization of an analogue of relativistic quantum mechanics in graphene[10-12] has led to much investigation[13,14] of the FQH effect in this system, in which combined spin and valley degrees of freedom provide new possible ground states[15], including multicomponent FQH states[16] with an unconventional sequence[17]. Numerous FQH states have been observed in the *N*=0 and *N*=1 LLs of monolayer graphene, but never in the *N*=2 LL (6 < |$v$| < 10) (Ref. 18). This is reminiscent of the situation in GaAs-based 2D electron systems, where in the *N*=2 LL (4 < $v$ < 6), Wigner crystal bubbles supplant the Laughlin state, although the 4+1/5 and 4+4/5 states have been observed[19]. A possible reason is that in high (*N*>0) LLs, the more extended wave-functions give rise to greater residual interactions between the CFs and may destabilize the FQH states[20].

With advances in sample preparation, the FQH effect was also recently seen in bilayer graphene[8,21-23]. Specific findings include tunability of states with electric field normal to the plane[21], even-denominator FQH states at $v$= -1/2 and at $v$= -5/2 (Ref. 22), and – in scanning compressibility measurements – particle-hole asymmetry in the *N*=0/*N*=1 LLs [8]. In the *N*=2 LL, incipient FQH states at $v$ = 14/3, 17/3, 20/3, and 23/3, which do not form a complete CF sequence, were seen[21]. In another transport study, particle-hole symmetric and asymmetric FQH states coexisted in the same measurement set[23]. Here, we report observations of particle-hole symmetric FQH states, consistent with a composite fermion accounting, in the $N=2$ LL in bilayer graphene.

Observation of the FQH effect requires ultra-clean systems whose disorder energy scale is less than the energy gaps of the elementary excitations from the fractional ground states[18]. We achieve the desired cleanliness by fabricating open-face bilayer-graphene/h-BN/graphite stacks sitting on $SiO_2$ (schematic in Fig. 1a and details in Supplementary Information). All the devices studied in this work were operated at densities ~1.5-5 x $10^{12}$ cm$^{-2}$ with zero-field mobility of 100,000-250,000 cm$^2$/V·s. Fig. 1b shows the typical strongly-insulating behaviour we observe near the charge neutrality point in these samples, with a width suggestive of low (~ $10^{10}$ cm$^{-2}$) disorder density[18] (Fig. 1b). Fig. 1c shows the longitudinal and Hall magnetoresistances ($R_{xx}$ and $R_{xy}$) in the $N\geq 2$ hole LL of device 1 as a function of magnetic field from 11.4 T to 45 T at $T$=0.4 K. The onset of broken symmetry occurs by 2 T (fig. 1c, inset) and fully symmetry-broken Hall plateaux are seen by 5 T (Supplementary Fig. 3). We also observe pronounced $R_{xx}$ minima at fractional filling $\nu$ = -13/3, -14/3, -16/3, and -17/3 in the $N=2$ LL, with accompanying plateau-like structures in $R_{xy}$ at $\nu$ = -13/3 and -14/3.

Fig. 2 provides more detailed data on the fractional states for $4 < |\nu| < 6$. When sweeping the back-gate voltage at $B$=30 T, we observe the $R_{xx}$ minima and the $R_{xy}$ plateaux of some of the fractional states (Fig. 2a). In addition to states with denominator 3, we also observe the more weakly-formed ±22/5, ±23/5, ±27/5, and ±28/5 FQH states (Fig. 2a and Fig. 2b). The -13/3 and -14/3 states, seen in device 1 at 30 T (Fig. 2a) have analogues at +13/3 and +14/3 at fields as low as 7 T in device 2 (Fig. 2d). A Landau fan diagram showing the -13/3 and -14/3 states is presented in Fig. 2c (device 3). In device 2, the states 19/3 and 20/3 (Fig. 2b) are the highest observed within-LL particle-hole symmetric pairs reported in any quantum Hall system; we do not observe any FQH states for ν > 7. Low-field $R_{xx}$ data as a function of filling factor (normalized carrier density) allow us to confirm the assignment of the fractional states to FQH features, which appear as vertical lines in Fig. 2d. Quantisation of the $R_{xy}$ plateau, when they are clearly discerned, is within 1% of $(1/\nu)(h/e^2)$. The FQH states we observe in the $N=2$ LL in bilayer graphene follow the composite fermion sequence with ν = $p/(2p\pm 1)$, yielding particle-hole conjugate states. Thus, the pairs ν= −13/3=-4+(-1/3) and ν = -14/3=-5-(-1/3); as well as ν = -16/3=-5+(-1/3) and ν = -17/3=-6+(-1/3), and similarly for the states with denominator 5, display particle-hole symmetry within each fully symmetry-broken LL. These results are significant and unexpected in light of recent experimental findings of particle-hole asymmetry in the FQHE in bilayer graphene.

Fig. 3a shows the strong temperature dependence of $R_{xx}$ for 4 < |ν|< 5 from device 3, on the hole side, which shows that the fractional states are largely suppressed above ~ 3 K. We used such data to extract the energy gap of four fractional states with denominator 3 (-13/3, -14/3) for -4 < ν < -5 (device 3) and 4 < ν < 6 (device 2), (Supplementary information) at several magnetic fields. The temperature-dependence of the $R_{xx}$ minima for the -13/3 and -14/3 states at 14 T fits the usual Arrhenius law $R_{xx} \sim \exp[-\Delta/(2T)]$, with $\Delta$ the FQH energy gap divided by the Boltzmann constant, and $T$ the temperature (fig. 3b). Based on the fits for the data at $B$=14 T, we calculate $\Delta_{-13/3}$ = (2.6±0.1) K (in units of Coulomb energy, ~0.01 $e^2/\varepsilon l_B$) and $\Delta_{-14/3}$ = (7.9±0.4) K (~0.03 $e^2/\varepsilon l_B$), and for the states at 16/3 and 17/3, $\Delta_{16/3}$ = (7.5±0.2) K (~0.032$e^2/\varepsilon l_B$) and $\Delta_{17/3}$=(7.0±0.6) K (~0.025$e^2/\varepsilon l_B$) where $l_B = (\hbar/eB)^{1/2}$ is the magnetic length.

Measured FQH gaps are normally significantly reduced by disorder broadening[24] and (secondarily) Landau level mixing[25]. In monolayer graphene[18] and the $N=1$ LL in GaAs/AlGaAs systems[26], FQH gaps are at least one order of magnitude smaller than values predicted in the absence of disorder. Assuming disorder affects the particle-hole conjugate states in the same way, the relative magnitude of the true activation gaps should track the measured ordering of gap sizes. Interestingly, while $\Delta_{16/3}$ is close in value to $\Delta_{17/3}$ across the magnetic field range used for the study of the activated gaps, $\Delta_{-14/3}$ is larger than $\Delta_{-13/3}$ across the field range; almost three times larger at 14 T. For states that follow the expected CF sequence, the activation gap of particle-hole conjugate states is expected to be the same, as seen in the single activation energy $^3\Delta$ measured for $\nu =1/3$ and 2/3, and $^5\Delta$ for 2/5 and 3/5 in GaAs[27]. The gap at -13/3 is smaller than that at -14/3: perhaps the different strength of electric field normal to the plane plays a role, or perhaps LL mixing is more important for -13/3 than -14/3. Finally, we may be close to a transition in quantum numbers of the partially-filled LL. In the Landau fan of Fig. 2c, the $\nu$=5 gap may be seen to weaken and then re-emerge as field is tuned through 9-10T ($V_g \sim$ -1.8V).

The FQH gaps we measure in $N$=2 LL are larger than those in the $N$=1/$N$=0 LLs, such as 5/3 ($\Delta \sim$ 2.5 K), 4/3 ($\Delta \sim$ 4 K), and 8/3 ($\Delta \sim$ 1.2 K) (Fig. S8). This runs counter to experimental work and predictions on conventional semiconductor QH systems whose FQH states are generally less stable than their counterparts in the lowest LL[4-9]. However, calculations on fractional states in bilayer graphene do predict robust thirds and fifths in the N=2 LL due to a shorter-range pseudopotential than in the N=2 LL in other systems[28].

Tilted magnetic-field measurements allow discrimination of the effects of Coulomb interactions (tuned by perpendicular field) from those of Zeeman splitting (tuned by total field). We compare $R_{xx}$ and $R_{xy}$ measured at a perpendicular field of 25 T with and without an in-plane component of ~37 T (Fig. S4). The -13/3, -14/3, -16/3 and -17/3 states show little $R_{xx}$ minima variation, suggesting that these states are spin-polarized at these fields.

Our results raise a fundamental question about whether the FQH states in bilayer graphene are described by the CF model in both the $N=0/N=1$ LLs and the $N=2$ LL. In the $N=0/N=1$ LLs, we observe fractional states that do not form a complete CF sequence. Fig. S5a shows $R_{xx}$ and $R_{xy}$ of device 4 at 30 T as a function of back-gate voltage throughout the accessible densities on the hole side. In this device, -7/3, -8/3 and -11/3 are seen in the $N=0/N=1$ LL, while 2/3, 4/3, 5/3 and 8/3 are seen in another device on the electron side (Fig. S5b), confirming the general observation from all measured samples that the states in the lowest LL do not form a complete CF sequence, unlike the observation in $N=2$ LL. The appearance of a full conventional CF sequence in $N=2$ LL also suggests that LL mixing, which should break particle-hole symmetry, is a modest perturbation rather than changing ground states in this LL.

Though relatively strong FQH states in the N=2 LL were anticipated in one early theoretical work on bilayer graphene[28], these states have proven even stronger than expected relative to those in the N=0/N=1 LL, warranting further investigation. The sharp contrast between our observations and prior studies of bilayer graphene may be related to differences between the heterostructures studied: notably, we designed our heterostructures to enhance Coulomb interactions (Ref. 18 and Supp. Mat.). This points to the opportunity to rationally optimize van der Waals heterostructures to host desired FQH states.

References:

1. Tsui, D. C., Stormer, H. L. & Gossard, A. C. Two-Dimensional Magnetotransport in the Extreme Quantum Limit. *Phys. Rev. Lett.* **48,** 1559–1562 (1982).
2. Laughlin, R. B. Anomalous Quantum Hall Effect: An Incompressible Quantum Fluid with Fractionally Charged Excitations. *Phys. Rev. Lett.* **50,** 1395–1398

(1983).

3. Jain, J. K. *Composite Fermions*. (2007). doi:10.1017/CBO9780511607561
4. Koulakov, A. A., Fogler, M. M. & Shklovskii, B. I. Charge Density Wave in Two-Dimensional Electron Liquid in Weak Magnetic Field. *Phys. Rev. Lett.* **76,** 499–502 (1996).
5. Töke, C., Peterson, M. R., Jeon, G. S. & Jain, J. K. Fractional quantum Hall effect in the second Landau level: The importance of inter-composite-fermion interaction. *Phys. Rev. B* **72,** 125315 (2005).
6. Fogler, M. M. & Koulakov, A. A. Laughlin liquid to charge-density-wave transition at high Landau levels. *Phys. Rev. B* **55,** 9326–9329 (1997).
7. Moessner, R. & Chalker, J. T. Exact results for interacting electrons in high Landau levels. *Phys. Rev. B* **54,** 5006–5015 (1996).
8. Kou, A. *et al.* Electron-hole asymmetric integer and fractional quantum Hall effect in bilayer graphene. *Science (80).* **345,** 55–57 (2014).
9. Girvin, S. M. Particle-hole symmetry in the anomalous quantum Hall effect. *Phys. Rev. B* **29,** 6012–6014 (1984).
10. Novoselov, K. S. *et al.* Electric field effect in atomically thin carbon films. *Science* **306,** 666–9 (2004).
11. Zhang, Y., Tan, Y.-W., Stormer, H. L. & Kim, P. Experimental observation of the quantum Hall effect and Berry's phase in graphene. *Nature* **438,** 201–204 (2005).
12. Novoselov, K. S. *et al.* Two-dimensional gas of massless Dirac fermions in graphene. *Nature* **438,** 197–200 (2005).
13. Bolotin, K. I., Ghahari, F., Shulman, M. D., Stormer, H. L. & Kim, P. Observation of the fractional quantum Hall effect in graphene. *Nature* **475,** 122 (2011).
14. Du, X., Skachko, I., Duerr, F., Luican, A. & Andrei, E. Y. Fractional quantum Hall effect and insulating phase of Dirac electrons in graphene. *Nature* **462,** 192–195 (2009).
15. Goerbig, M. O. & Regnault, N. Analysis of a SU(4) generalization of Halperin's wave function as an approach towards a SU(4) fractional quantum Hall effect in graphene sheets. *Phys. Rev. B* **75,** 241405 (2007).
16. Dean, C. R. *et al.* Multicomponent fractional quantum Hall effect in graphene. *Nat Phys* **7,** 693–696 (2011).
17. Feldman, B. E., Krauss, B., Smet, J. H. & Yacoby, A. Unconventional Sequence of Fractional Quantum Hall States in Suspended Graphene. *Science (80).* **337,** 1196–1199 (2012).
18. Amet, F. *et al.* Composite fermions and broken symmetries in graphene. *Nat*

*Commun* **6,** 5838 (2015).

19. Gervais, G. *et al.* Competition between a Fractional Quantum Hall Liquid and Bubble and Wigner Crystal Phases in the Third Landau Level. *Phys. Rev. Lett.* **93,** 266804 (2004).
20. Xia, J. S. *et al.* Electron correlation in the second Landau level: a competition between many nearly degenerate quantum phases. *Phys. Rev. Lett.* **93,** 176809 (2004).
21. Maher, P. *et al.* Tunable fractional quantum Hall phases in bilayer graphene. *Science (80).* **345,** 61–64 (2014).
22. Ki, D.-K., Fal'ko, V. I., Abanin, D. A. & Morpurgo, A. F. Observation of Even Denominator Fractional Quantum Hall Effect in Suspended Bilayer Graphene. *Nano Lett.* **14,** 2135–2139 (2014).
23. Kim, Y. *et al.* Fractional Quantum Hall States in Bilayer Graphene Probed by Transconductance Fluctuations. *Nano Lett.* **15,** 7445–51 (2015).
24. Pan, W. *et al.* Experimental studies of the fractional quantum Hall effect in the first excited Landau level. *Phys. Rev. B* **77,** 075307 (2008).
25. Willett, R. L., Stormer, H. L., Tsui, D. C., Gossard, A. C. & English, J. H. Quantitative experimental test for the theoretical gap energies in the fractional quantum Hall effect. *Phys. Rev. B* **37,** 8476–8479 (1988).
26. Choi, H. C., Kang, W., Das Sarma, S., Pfeiffer, L. N. & West, K. W. Activation gaps of fractional quantum Hall effect in the second Landau level. *Phys. Rev. B* **77,** 081301 (2008).
27. Boebinger, G. S. *et al.* Activation energies and localization in the fractional quantum Hall effect. *Phys. Rev. B* **36,** 7919–7929 (1987).
28. Shibata, N. & Nomura, K. Fractional Quantum Hall Effects in Graphene and Its Bilayer. *J. Phys. Soc. Japan* **78,** 104708 (2009).

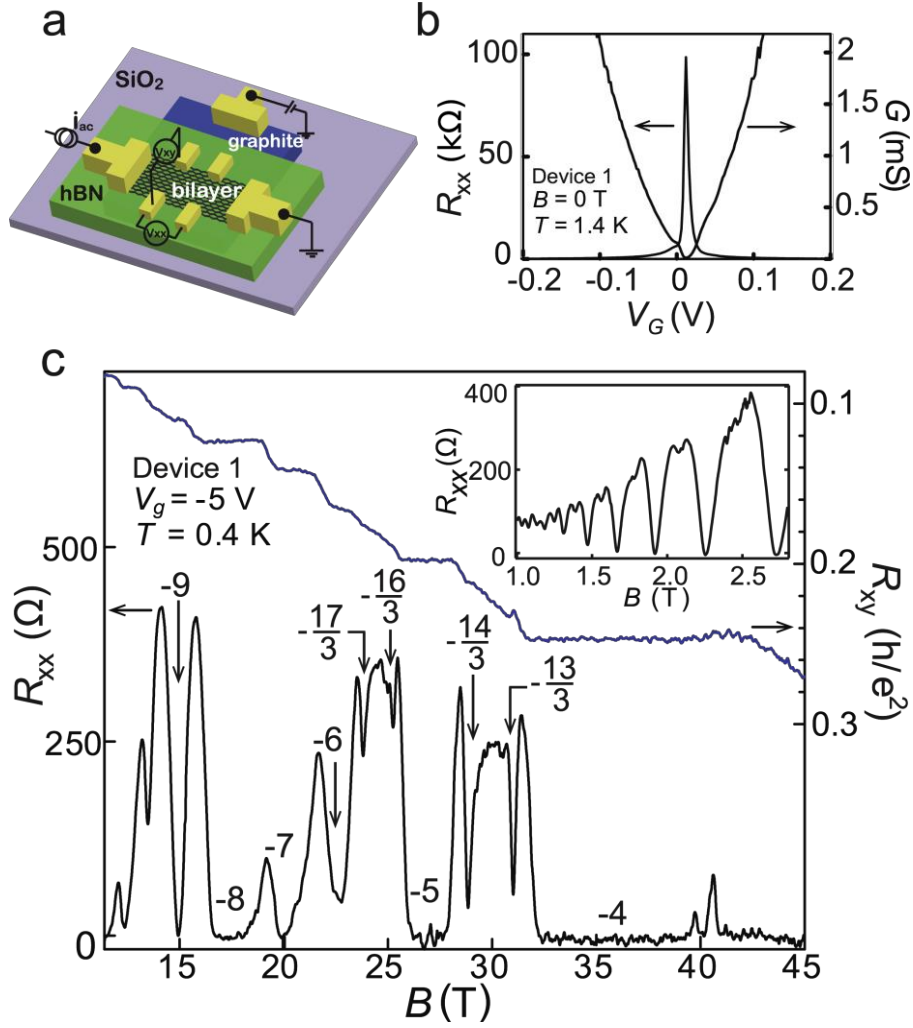

**Figure 1. Device schematic and transport characteristics.** **a.** Schematic of our bilayer graphene device design. **b.** Zero-field resistance $R_{xx}$ and its inverse, conductance $G$, as a function of graphite back gate voltage $V_g$ for device 1 (optical image in Fig. S1). **c.** Magnetoresistance $R_{xx}$ and Hall resistance $R_{xy}$ as a function of magnetic field $B$ at $V_g$ = -5 V. Corresponding Landau level filling factors are labelled. Inset: low-field magnetoresistance with Shubnikov-de Haas oscillations showing the onset of degeneracy breaking among the integer states.

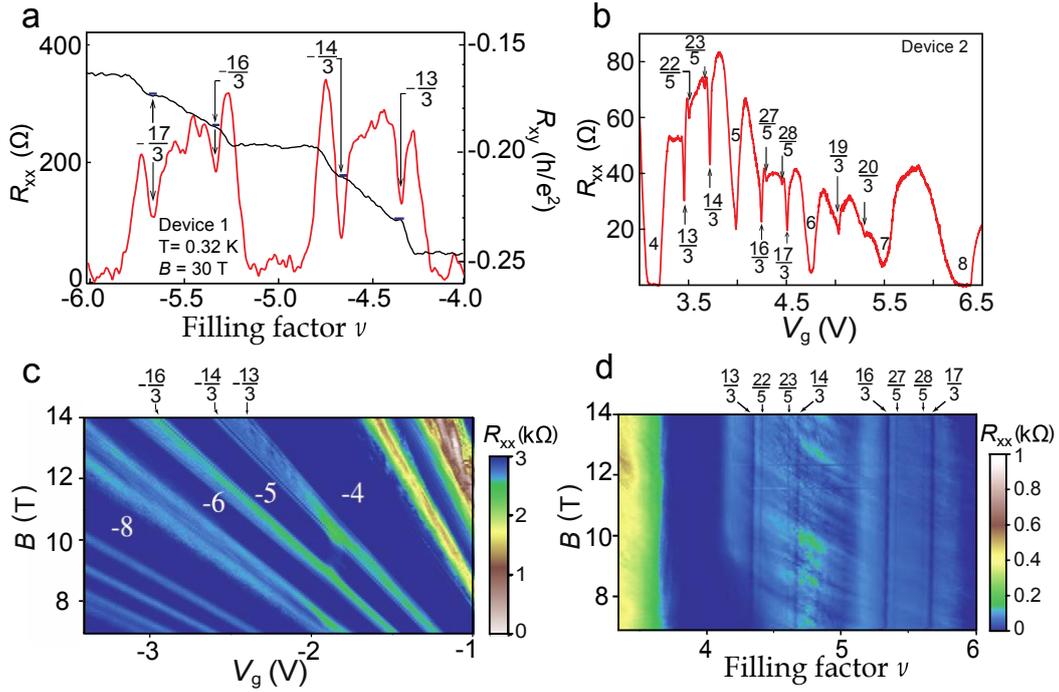

**Figure 2. Particle-hole symmetric fractional quantum Hall effect in the $N=2$ Landau level. a.** Magnetoresistance $R_{xx}$ and Hall resistance $R_{xy}$ of device 1 at 30 T showing pronounced fractional states with denominator 3 and more weakly-developed states with denominator 5. **b.** The same fractional states seen in $R_{xx}$ on the electron side of device 2 at 14 T. **c.** Landau fan diagram of $R_{xx}$ as a function of magnetic field and carrier density on the hole side for device 3. **d.** $R_{xx}$ as a function of filling factor (carrier density rescaled by magnetic field) for device 2 on the electron side. Vertical features mark FQH states.

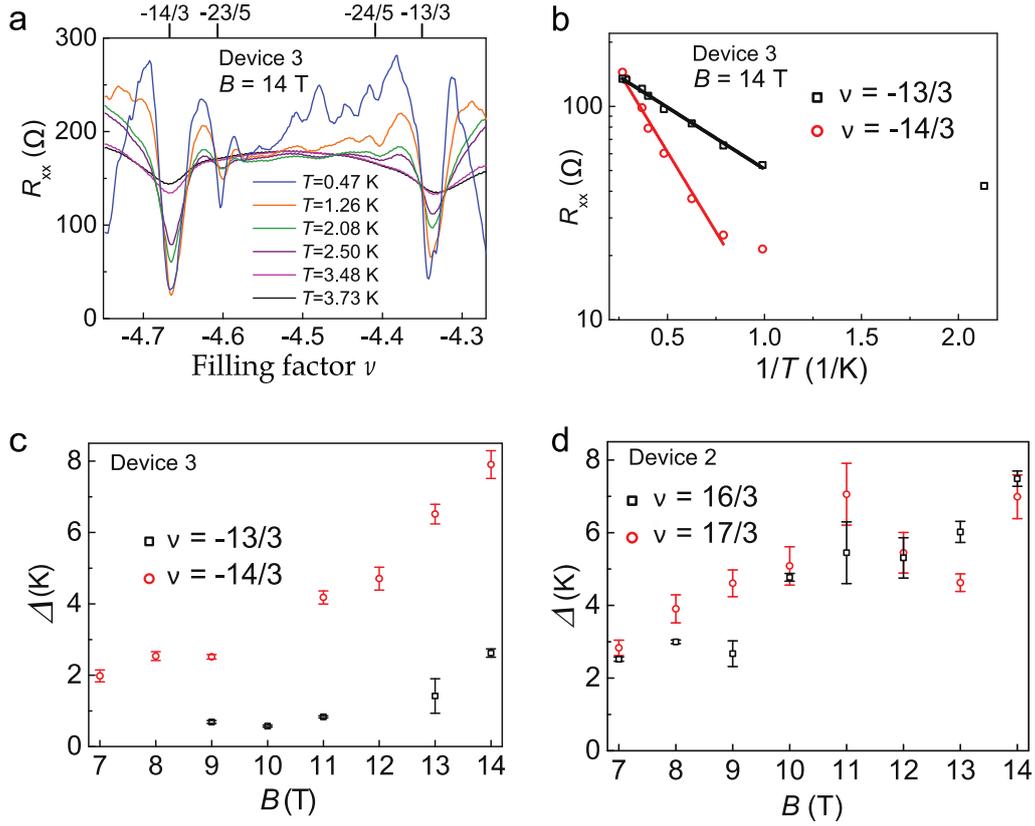

**Figure 3. Fractional quantum Hall gaps in the *N*=2 Landau level. a**. Temperature dependence of the magnetoresistance $R_{xx}$ for device 3 between ν=−5 and -4 at 14 T, showing that the $R_{xx}$ minima for the states with denominator 3 deepen with decreasing temperature down to ~ 1K. **b.** $R_{xx}$ at ν= -13/3 and -14/3 (device 3) plotted on a semi-logarithmic scale as a function of inverse temperature. The linear fits yield activation gaps, greater for ν=-14/3. The lowest-temperature data points depart from activated behavior, as is typically seen in QH systems at the onset of variable-range hopping and stronger localization. **c**. Measured gaps as a function of magnetic field for ν=-13/3 and -14/3 (Device 3) and **d.** for ν=16/3 and 17/3 (electron side in device 2).

**Acknowledgments**

We thank Eli Fox for experimental help. We thank Michael Zaletel, Zlatko Papic, Michael Peterson and Kiryl Pakrouski for theoretical discussions about FQH in higher LLs, and for sharing unpublished calculations. Luis Balicas graciously commented on an earlier version of the manuscript. A part of the measurements were performed at National High Magnetic Field Laboratory, which is supported by the US National Science Foundation cooperative agreement no. DMR-1157490. Experiments were funded in part by the Gordon and Betty Moore Foundation through Grant GBMF3429 to D.G-G. G.D. was supported partly by a clean-energy seed grant from the Precourt Institute at Stanford University. P.G. acknowledges a Stanford Graduate Fellowship. M.L. acknowledges support from Samsung and Stanford University. A.J.B. was supported by a Benchmark Stanford Graduate Fellowship. C.T.L. was funded by the MOST, Taiwan (grant numbers MOST 103-2918-I-002-028 and MOST 103-2622-E-002 -031).

**Methods**

The bilayer graphene devices were made by the procedure described in Ref. 18 except that as-exfoliated bilayer graphene was chosen instead of monolayer graphene. We made no attempt to rotationally align our bilayer with h-BN flakes[18]. The experiments were performed in a cryogen-free dilution refrigerator and in a $^3$He cryostat using standard ac lock-in techniques. Measurements at fields higher than 14T were performed at the National High Magnetic Field Laboratory in Tallahassee, USA.

**Author contributions**

G.D., F.A. and D.G-G. conceived the experiment; G.D., P.G., and F.A. designed and

**Supplementary Information**

**1. Fabrication details**

We fabricated open-face bilayer graphene samples sitting on atomically-smooth hexagonal boron nitride (h-BN) layers. Briefly, suitably uniform, thin (~ 5-10 nm) graphite sheet exfoliated on a $SiO_2/n^{++}$ Si substrate were chosen to serve as a local bottom gate. Hexagonal boron nitride sheets were separately exfoliated on thin (~ 60 nm) PVA films deposited on bare Si with spin-coating. Suitable h-BN flakes were chosen using optical and AFM imaging, and were subsequently transferred onto a part of the bottom-gate graphite sheets using PMMA transfer and then annealed in 10% $O_2$ in Ar at 500° C [Ref. 1]. Bilayer graphene was then transferred on the h-BN sheets and Hall bars were fabricated. We made no attempt to rotationally align our bilayers with the underlying h-BN flakes and we saw no signs of secondary (superlattice) Dirac peaks and fractal quantum Hall states in our devices[2,3]. For $B > 1.5$ T, the magnetic length $l_B = 26/\sqrt{B}$ (in nm) is always shorter than the h-BN thickness (27 nm or 47 nm) so that the graphite back-gate suppresses potential fluctuations without screening the short-range Coulomb interactions responsible for the FQH states. Fabrication of such open-face samples was successfully accomplished with both PMMA (wet) transfer and PDMS-PPC (dry) transfer.

For device 1 (Fig. S1), the calculated field-effect mobility (the slope of the conductance) was ~200,000 $cm^2$ $V^{-1}s^{-1}$ at density $10^{11} cm^{-2}$. The devices we measured were strongly insulating at the charge neutrality point at low temperatures, with resistivity above 200 k$\Omega$ in some devices at $T$= 4 K.

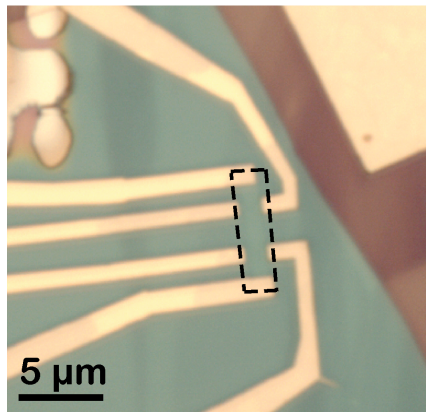

**Figure S1.** Optical image of device 1.

The decision to fabricate open-face samples, rather than the recently reported encapsulated fabrication method[4,5], was deliberate. Fully encapsulating graphene in atomically-smooth h-BN layers has led to exceptional cleanliness, as reflected in mobility and mean-free path[4]. Encapsulated devices consist of an h-BN/graphene/h-BN sandwich, with possible additional bottom and top gate (each of which may be graphite or metal). An encapsulated structure with a top metallic gate was recently used in the study of electric field-induced transitions between FQH states in bilayer graphene[5]. Encapsulation requires the formation of one-dimensional metal contacts to the edges of the graphene mesa. Such contacts have proven remarkably robust at low magnetic field, but the situation is more complicated at high field: the deposited metal of such contacts by necessity extends at least a small distance onto the top of the h-BN layer, above the encapsulated graphene. The graphene sheet is thus exposed to different electrostatic environments in different areas, leading to non-uniform carrier density and, upon application of a perpendicular magnetic field, different filling factors. Underneath the metal contact, the carrier density (with magnetic field, filling factor $\nu_1$) will be determined in part by the differences in the work functions between graphene and the contact metal. A different carrier density (different filling factor, $\nu_3$) will form in the interior of the sample as a result of the electrostatic gating. Between these two areas, a small area of transitional density (filling factor $\nu_2$) will form. Contacts may be improved by applying a large voltage to the doped Si substrate, to heavily dope the graphene in the contact region with the same polarity as bulk of the film (see Supplementary Material of Ref. 5), but the presence of multiple different filling factors may remain an important complication.

In comparison, we plot a schematic diagram of the open-face graphene devices that we fabricated for the FQH measurements. In the region under the ohmic contact, the filling factor ($\nu_1$) will be different from that in the main region of the device ($\nu_3$). However, since the region underneath the metal contact is highly disordered and can be regarded as a reservoir for carriers, quantum Hall measurements should not be affected. Although there could exist a region with an intermediate filling factor ($\nu_2$) between the dashed lines in an open-face device, the width of this region would be substantially shorter than that in an encapsulated device. Moreover, we note that the Coulomb energy of the FQH states is given by $e^2/\varepsilon l_B$. In our open-face devices, the effective dielectric constant is given by the average of the dielectric constants of h-BN and vacuum, $\varepsilon = (4+1)/2=2.5$, which is smaller compared with that in an encapsulated device, where $\varepsilon = 4$. This reduction in the effective dielectric constant favors larger FQH gaps in graphene.

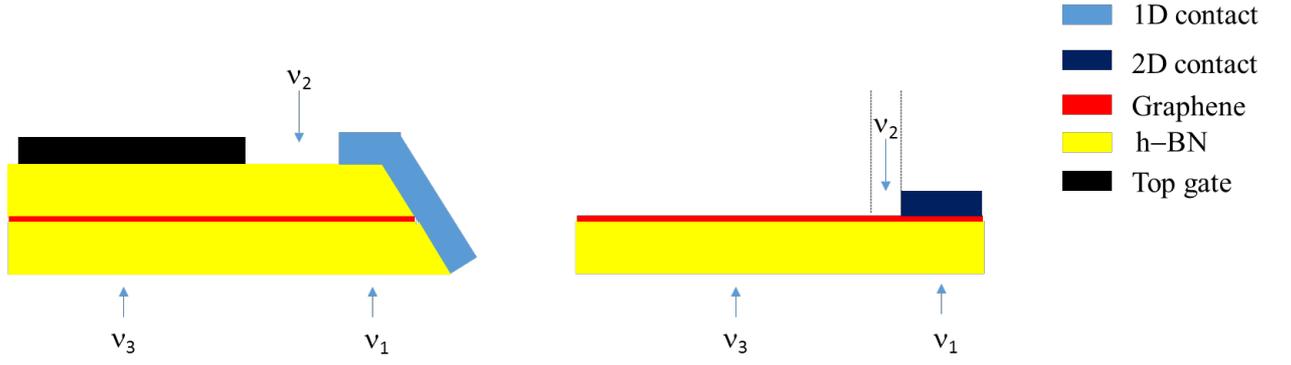

**Figure S2.** Left panel: A schematic diagram showing regions with different filling factors in a h-BN/graphene/h-BN structure with a one-dimensional contact. Contact-induced doping will make the filling factor near the contact region ($\nu_1$) different from that in the interior that is gated ($\nu_3$) and the intermediate region ($\nu_2$). Right panel: A schematic diagram showing regions with different filing factors in an open-face bilayer graphene device. The graphene region underneath the Ohmic contact is at filling factor $\nu_1$. There exists a small region between the dashed lines at filling factor $\nu_2$. The main region of the device is at filling factor $\nu_3$

## 2. Integer quantum Hall states in device 1.

A full set of symmetry-broken integer quantum Hall states can be observed at integer filling factors for $B > 5$ T.

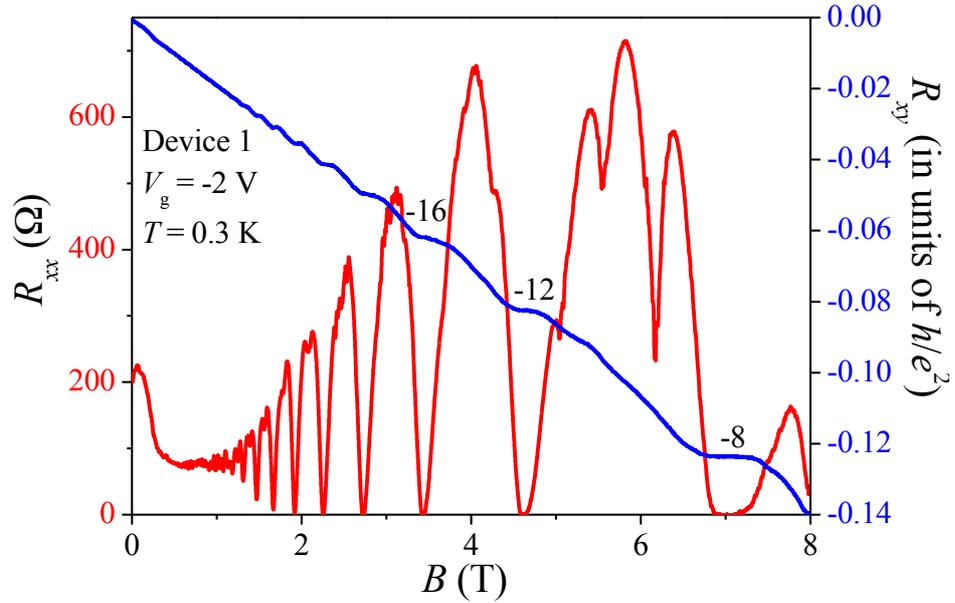

**Figure S3**. Low-field magnetoresistance measurements on device 1.

## 3. Tilted-field measurements on device 1.

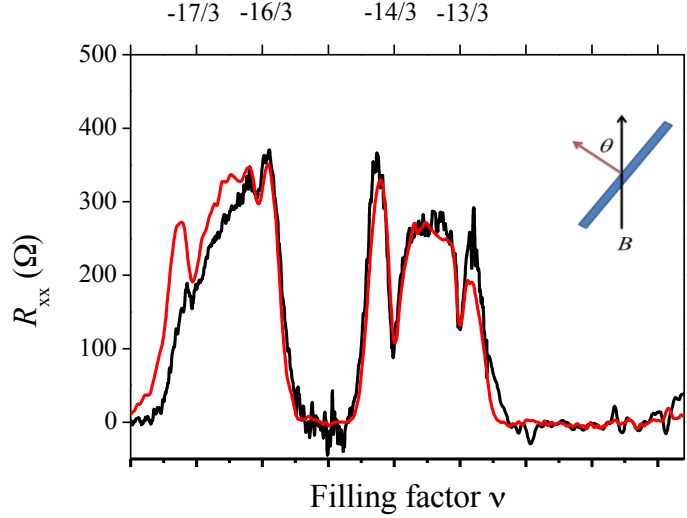

**Figure S4.** Tilted-field measurement for the thirds in the region $4 < |\nu| < 6$. The red curve corresponds to $B_{Tot} = B_\perp = 25$ T. The black curve corresponds to a tilt angle of $\theta = 56.3°$, leading to $B_{Tot} = 45$ T and $B_\perp = 25$ T).

The tilted-field results in Fig. S4 suggest that the ground states of the thirds seen in the range $4 < |\nu| < 6$ are spin-polarized. At both angles $\theta = 0°$ and $\theta = 56.3°$, the perpendicular magnetic field ($B_\perp$), which sets the energy scale of the Coulomb interaction, is fixed at 25 T. Since the scale of the Zeeman energy is governed by the total magnetic field, which changes from 25 T to 45 T, changes in the fractional states would be indicative of the ground-state being spin-unpolarized[6]. With the possible exception of -17/3, which may be slightly weakened, the fractional states are not impacted by the increased Zeeman energy.

## 4. FQH states in the N=0/N=1 LLs

a

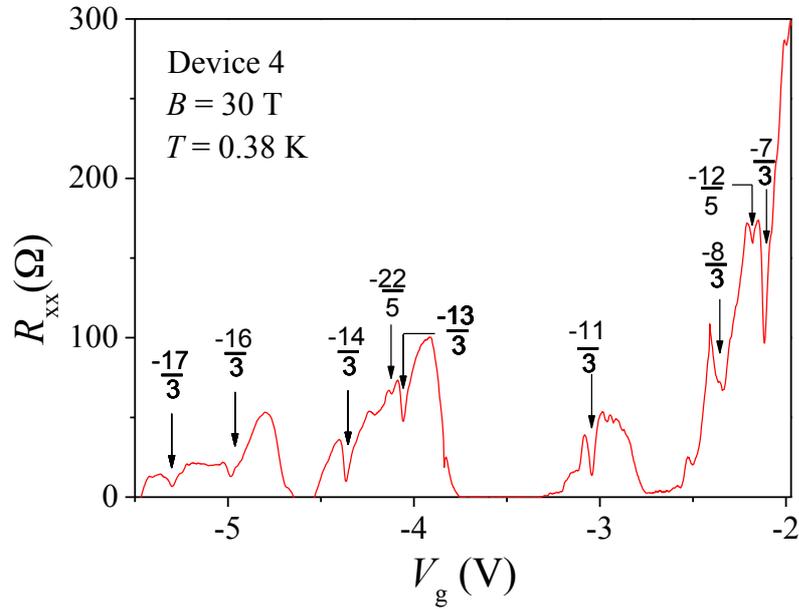

b

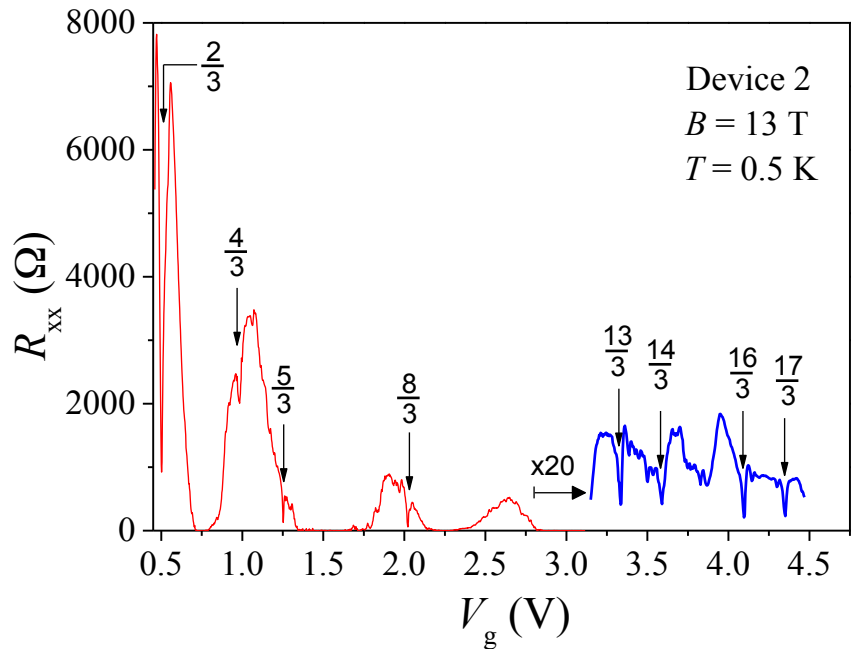

**Figure S5a and S5b.** Magnetoresistance data showing FQH states in the $N=0/N=1$ LLs and the $N=2$ LL.

## 5. Resonant structures and temperature dependence of $R_{xx}$ for $4 < \nu < 6$ in device 2

In device 2 the presence of resonant features for $4 < \nu < 5$, as seen in the Landau fan diagram in Fig. 2d and in Fig. S6, may prevent accurate quantification of the FQH gaps.

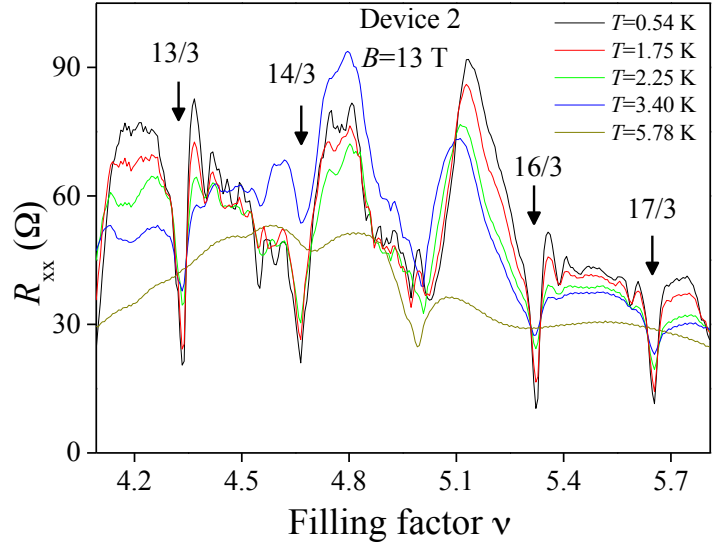

**Figure S6.** $R_{xx}$ as a function of filling factor for $B=13$ T at various temperatures (device 2).

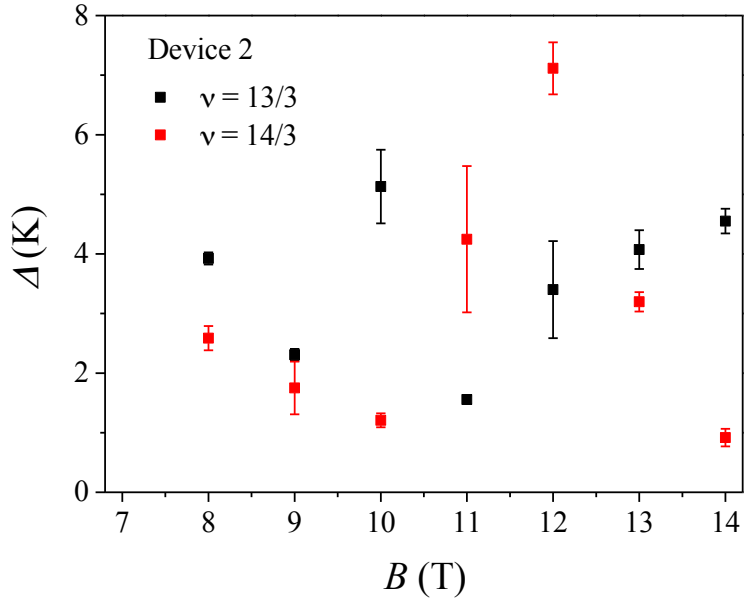

**Figure S7.** Measured gap in Kelvin as a function of magnetic field for $\nu=13/3$ and $\nu=14/3$ (device 2).

## 6. Magnetic field dependence of activation gaps.

The magnetic-field dependence of the energy gaps of the thirds we have observed in the range of 4 < |ν|< 6 is shown in fig. 3c-d and Fig. S7. The magnitude of the energy gaps of spinless quasiparticle excitations is expected to vary with field as $\sqrt{B}$. The gaps we measure generally increase with field, though we do not follow them over a sufficient field range to extract a particular power law. The measurable gaps of the four fractional states on which we focus exhibit a linear or slightly superlinear temperature dependence.

## 7. Overview of gaps in the $N=0/N=1$ LLs and $N=2$ LL in device 2

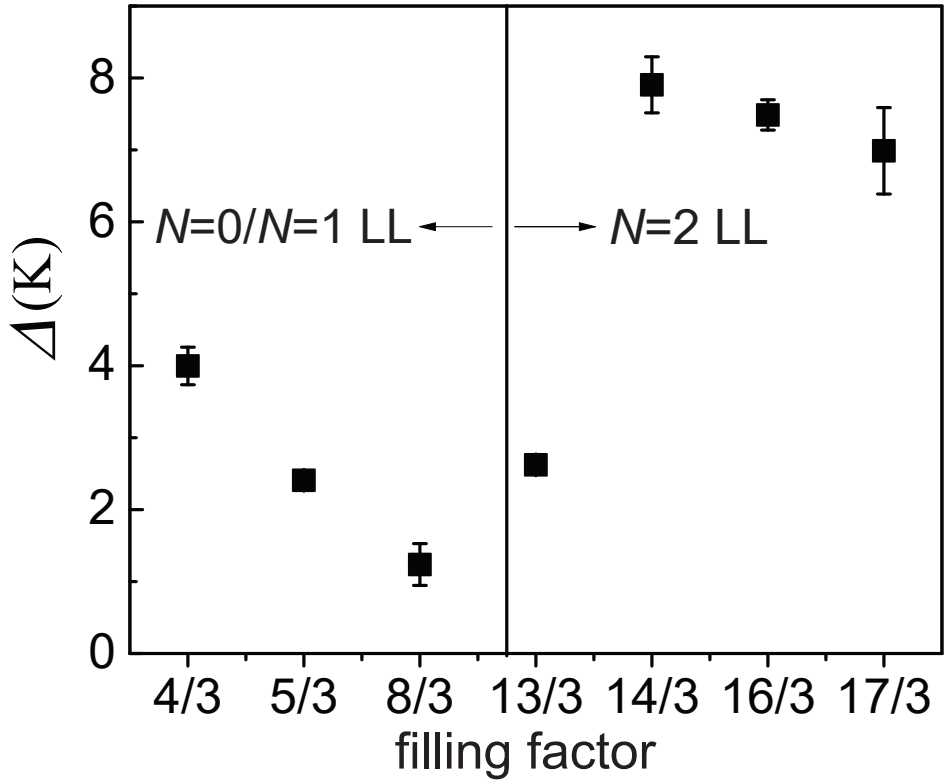

**Figure S8.** The measured activation gaps of the thirds seen in the $N=0/N=1$ LLs and $N=2$ LL (device 2).

The gaps in the $N=2$ LL are approximately twice as large as those in the $N=0/N=1$ LLs. This behaviour is opposite to that seen in GaAs, where FQH states in the $N=2$ LL are fragile and experiments at extremely low temperatures are needed to extract meaningful activation gaps[7]. The inverted scale of gaps between $N=0$ and higher LLs is also seen in monolayer graphene, where robust states (which, however, do not form a complete CF sequence) are seen in the $N=1$ LL (Ref. 6).

References

1. Garcia, A. G. F., Neumann, M., Amet, F., Williams, J. R., Watanabe, K., Taniguchi, T., & Goldhaber-Gordon, D. Effective cleaning of hexagonal boron nitride for graphene devices. *Nano Lett.* **12**, 4449−4454 (2012).

2. Yankowitz, M., Xue J., Cormode D., Sanchez-Yamagishi, J. D., Watanabe K., Taniguchi T., Jarillo-Herrero P., Jacquod P. & LeRoy B. J. Emergence of superlattice Dirac points in graphene on hexagonal boron nitride. *Nature Phys.* **8**, 382-386 (2012).

3. Ponomarenko, L. A., Gorbachev, R. V., Yu, G. L., Elias, D. C., Jalil, R., Patel, A. A., Mishchenko, A., Mayorov, A. S. Woods, C. R., Wallbank, J. R., Mucha-Kruczynski, M., Piot, B. A., Potemski, M., Grigorieva, I. V., Novoselov, K. S., Guinea, F., Fal'ko V. I., & Geim, A. K. *Nature* **497**, 594-597 (2013).

4. Wang, L., Meric, I., Huang, P. Y., Gao, Q., Gao, Y., Tran, H., Taniguchi, T., Watanabe, K., Campos, L. M., Muller, D. A., Guo, J., Kim, P., Hone, J., Shepard, K. L. & Dean, C. R. One-dimensional electrical contact to a two-dimensional material. *Science* **342**, 614-617 (2013).

5. Maher, P., Wang, L., Gao, Y., Forsythe, C., Taniguchi, T., Watanabe, K., Abanin, D., Zlatko Papić, Z. Cadden-Zimansky, P., Hone, J., Kim, & P. Dean, C. R. Tunable fractional quantum Hall phases in bilayer graphene. *Science* **345**, 61-64 (2014).

6. Amet, F., Bestwick, A. J., Williams, J. R., Balicas, L., Watanabe, K., Taniguchi, T., & Goldhaber-Gordon, D. Composite fermions and broken symmetries in graphene. *Nat Commun* **6,** 5838 (2015).

7. Gervais, G., Engel, L. W., Stormer, H. L., Tsui, D. C., Baldwin, K. W., West, K. W., & Pfeiffer L. N. Competition between a fractional quantum Hall liquid and bubble and Wigner crystal phases in the third Landau level. *Phys. Rev. Lett.* **93,** 266804 (2004).